# The Price of Tailoring the Index to Your Data: Poisoning Attacks on Learned Index Structures

Evgenios M. Kornaropoulos
George Mason University, USA
evgenios@gmu.edu

Silei Ren
Cornell University, USA
sr2262@cornell.edu

Roberto Tamassia
Brown University, USA
roberto@tamassia.net

## ABSTRACT

The concept of *learned index structures* relies on the idea that the input-output functionality of a database index can be viewed as a prediction task and, thus, implemented using a machine learning model instead of traditional algorithmic techniques. This novel angle for a decades-old problem has inspired exciting results at the intersection of machine learning and data structures. However, the advantage of learned index structures, i.e., the ability to adjust to the data at hand via the underlying ML-model, can become a disadvantage from a security perspective as it could be exploited.

In this work, we present the first study of data poisoning attacks on learned index structures. Our poisoning approach is different from all previous works since the model under attack is trained on a cumulative distribution function (CDF) and, thus, every injection on the training set has a cascading impact on multiple data values. We formulate the first poisoning attacks on linear regression models trained on a CDF, which is a basic building block of the proposed learned index structures. We generalize our poisoning techniques to attack the advanced two-stage design of learned index structures called recursive model index (RMI), which has been shown to outperform traditional B-Trees. We evaluate our attacks under a variety of parameterizations of the model and show that the error of the RMI increases up to 300× and the error of its second-stage models increases up to 3000×.

## 1 INTRODUCTION

Database systems rely on *index structures* to access stored data efficiently. It is known to the database community that the motto "one size fits all" does not apply to traditional indexing schemes [24] since each index provides different performance guarantees that depend on the access pattern, the nature of the workload, and the underlying hardware. Even after choosing an appropriate index structure for a specific application, it is usually the case that a database administrator has to manually fine-tune the parameters of the system, either through experience or with help from tools. The work by Kraska, Beutel, Chi, Dean, and Polyzotis [30] challenged the state of affairs by re-framing index structures as a *machine learning* problem where the index directs a query to a memory location(s) based on a trained model tailored on the data at hand.

**Learned Index Structures.** The core idea of a *learned index structure* (*LIS*) is to model a data structure as a *prediction task*, i.e., get an input key and predict its location in a sorted sequence of key-record pairs. This approach allows the use of (*i*) continuous functions to encode the data, and (*ii*) *learning algorithms* to approximate the function. The specific LIS approach proposed by Kraska *et al.* [30] is to build the *cumulative distribution function* (CDF) for the keys. Given a key, $k$, the CDF returns the probability that a key chosen according to this distribution takes value less than or equal to $k$. Since the above probability is built from the set of keys at hand, it is expressed as the ratio of the number of keys less than $k$ to the total number of keys. Given this insight, one can use the CDF to (i) compute the number of keys less than the (queried) key $k$, and (ii) infer the key's memory location assuming the keys were sorted during the initialization. Therefore, a simple linear regression on the CDF gives an approximate location of the queried key. Indeed a linear regression on the CDF is one of the building blocks that has been shown to work well [30] and can be combined with *hierarchical models*, also called recursive model index (RMI) structures, so as to balance the final model for latency, memory usage, and computational cost. The hierarchy can be seen as building a mixture of "experts" [39] responsible for subsets of the data. The notion of a LIS has spurred a surge of works that blend ideas from machine learning, data structures, and systems (e.g., [5, 7, 10, 12–14, 17–21, 23, 24, 29, 31, 35, 36, 41, 43, 45, 47–49, 52, 55, 56, 58, 59]).

**First Vulnerability Assessment of Learned Index.** As promising as it may sound to combine ideas from machine learning and data structures, no analysis has been performed to understand potential vulnerabilities of the LIS paradigm. Intuitively, the advantage of a LIS is that the model adapts to the data at hand. However this efficiency might be problematic if the adversary is capable of injecting *maliciously crafted data* before the training of the model, i.e., at the initialization stage of the index structure, so as to cause inaccurate predictions of the location of legitimate data.

The technique of *data poisoning* has been known to be an effective attack vector for over a decade, e.g., see the references in [26]. In the context of static index structures, we focus on the case where the data stored in the index comes from multiple sources as different entities directly or indirectly contribute data, e.g., by generating data with their actions or behavior. A malicious actor can tailor its contributed data to deteriorate the index performance. Indeed, the real-world datasets used in the original LIS work [30] come from multiple contributors and, thus, are susceptible to poisoning attacks. Other examples of indexed data generated by multiple sources include data from personalized medicine, where patients voluntarily contribute their own data, as well as cybersecurity analytics where any user can submit its own indicators of compromise. Our threat model, much like all poisoning works [3, 4, 26, 60, 61], assumes that the adversary has the ability to contribute maliciously crafted data and has access to the legitimate (training) data.

We analyze this blind spot by addressing the following question:

> Is it possible to **degrade the performance** of learned index structures by poisoning its learned components?

A first attempt to answer this question is to directly deploy previous poisoning attacks in this new CDF setting. Unfortunately, none of the proposed poisoning attacks applies to models based on a CDF. This is because all previous attacks operate under the assumption that any point $(X, Y)$ chosen by the adversary is a valid poisoning point, i.e., $X$-, $Y$-value can take any *arbitrary* number. On the contrary, when dealing with a point $(X, Y)$ from a CDF, the dimension $Y$ depends on the $X$-values of the *entire data set* since $Y$ describes the rank of its associated key. Thus, $Y$ can only take a *single* value given an $X$-value. This new and restrictive CDF setting forbids the use of any previous attacks and signals that we need a new methodology for addressing data poisoning attacks on CDFs.

**Our Contributions.** The purpose of this work is to ascertain whether the paradigm of learned systems is vulnerable to poisoning attacks. Towards this end we make the following contributions:

- We formulate the problem of *single-point poisoning* a linear regression model on the cumulative distribution function (CDF). We propose a poisoning attack algorithm on CDFs that finds a near-optimal poisoning key and has complexity linear to the number of legitimate keys.
- We propose a *multiple-point poisoning* attack on linear regression on CDF. We follow a greedy approach based on our efficient single-point attack. We conduct extensive experiments under a variety of scenarios to study the impact of different parameterizations. Our experiments show that the effect of our multi-point poisoning attack can increase the mean squared error (typical loss function of regression) up to 100×. To fix this accuracy error, the system may have to access up to 10× more memory locations.
- We propose a poisoning attack for the two-stage recursive model index (RMI) architecture that was shown to outperform [30] the highly-optimized traditional B-Tree data structure. Our attack targets the linear regression models of the second stage of the RMI. We evaluate the attack in uniformly and highly skewed key distributions under a variety of RMI models. We also apply the attack on real data from (i) the salaries of employees in Dade County in Miami [38], and (ii) publicly available geolocation dataset [44]. Our experiments show that the poisoning attack can increase the error of the RMI model up to 300× and the error of an individual second-stage model up to 3000×.

We hope our findings will motivate a systematic exploration of the robustness of learned systems by researchers in data management, machine learning, security, and systems.

## 2 RELATED WORK

Since the first work on learned index structures [30] the community has explored several interesting research directions [19, 58]. In terms of *dynamizing the LIS framework*, the work by Hadian and Heinis [18] proposes a method to minimize the error caused by updating the index. The first family of updatable LIS is called ALEX [7]. The work by Tang *et al.* [52] studies the case where the *distribution of the workload evolves* as queries are issued and proposes to re-train the model to account for such a dynamic environment. Another recent interesting line of work proposes learned *multi-dimensional* structures [8, 35, 43]. Pavo [59] takes an unsupervised neural network approach, for locating the position of the key value in the index. The combination of ML and data structures from [30] has inspired new ML-based approaches for traditionally algorithmic problems such as low-rank decomposition [25], space partitioning for nearest neighbor search [9], frequency estimation in data streams [21]. The work by Setiawan *et al.* [47] observes that the underlying process of LIS can be seen as a fitting problem, as opposed to a learning problem. The authors deploy polynomial interpolation techniques to approximate the location of a key. The PGM INDEX [15] proposes a model that adopts an RMI design that auto-tunes over space and latency requirements and supports updates. Based the PGM index, RADIXSPLINE [28] proposes a RMI that features an alternate linear interpolation based indexing. RadixSpline is efficient to build and tune, but falls back to tree-structured radix table on heavily skewed dataset.

System designs are also inspired by the work in [30]. SAGEDB [29] is a database that adapts to an application through code synthesis and ML. DECIMA [36] deploys reinforcement learning and ML to learn workload-specific scheduling algorithms. A line of works in benchmarking [27, 37] is also proposed to show the effectiveness of Learned Index Structures over traditional structures.

For completeness we note that other works considered a similar approach [2, 23, 33, 53, 54].

**Known Poisoning Attacks.** The line of work for *poisoning attacks* [4, 60, 61] considers adversaries that deliberately augment the training data to manipulate the results of the predictive model. The work by Biggio *et al.* [3] inserts maliciously crafted training data to change the decision function of the SVM and, thus, increase the test error. Yang *et al.* [61] propose gradient-based methods for generating poisoning points for Neural Networks. Suciu *et al.* [50] propose a framework for evaluating *realistic adversaries* that conduct poisoning attacks on machine learning algorithms. Finally, the work that is closest to ours is the results by Jagielski *et al.* [26]. The authors propose an optimization framework for poisoning attacks on linear regression as well as a defense mechanism called TRIM. However, we emphasize that our attacks differ significantly from [26] since we address the case of poisoning *CDFs* for which every insertion affects the values of *all points of the dataset*.

## 3 PRELIMINARIES

**Terminology.** We denote a key by $k$ and its key universe as $\mathcal{K}$, where $|\mathcal{K}| = m$. The set of all keys of an index is denoted as $K \subseteq \mathcal{K}$. The set of keys $K$ has size $n$ and contains no multiplicities. The *density* of a keyset $K$ is the ratio $|K|/|\mathcal{K}| = n/m$. For simplicity, we assume that keys are non-negative integers, therefore, we can always derive the total order of the keyset, much like [30]. Each key is associated with a record; the records are stored at an in-memory dense array that is sorted with respect to the key values.

### 3.1 Background on Learned Index Structures

The work by Kraska *et al.* [30] proposes alternative ML-based implementation for index structures such as *range indexes*, traditionally implemented by B-trees, *point indexes*, traditionally implemented by HashMap, and *existence indexes*, traditionally implemented by Bloom filters. Learned Index Structures (LIS) are based on a simple yet powerful observation that locating a key $k$ within a set of linearly ordered keys can be reduced to approximating the probability that a random key would take value less or equal than $k$, i.e., $\Pr(X \leq k) = \textsc{Rank}(k)/n$, where $X$ is the random variable that follows the empirical distribution of the $n$ keys and $\textsc{Rank}(\cdot)$ is a function that takes a key as an input and outputs its relative position among the $n$ keys. This probability is captured by the cumulative distribution function (CDF), thus, the task of locating a key boils down to *learning the CDF of the sorted key set*. We consider the non-normalized CDF, therefore, in a CDF plot the $X$-axis represents the key values and the $Y$-axis the rank of the key, see Figure 1. For generalizing to complex distributions the authors propose the *Recursive Model Index* (RMI), a multi-stage architecture where models on a higher stage direct the query to models on a lower stage to fine-tune the precision of the predicted memory location.

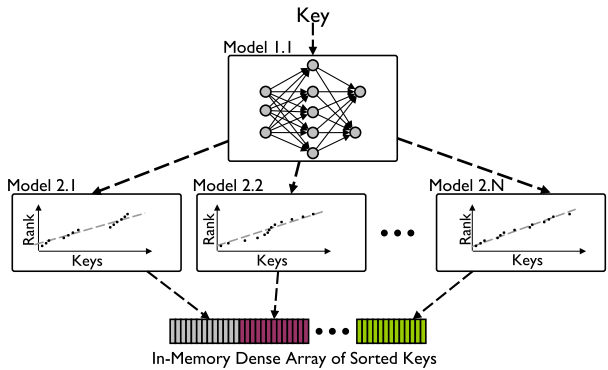

**Figure 1: An illustration of the Recursive Model Index (RMI) with a two-stage architecture. The first stage is a single neural network model while the second stage is series of linear regression models on 1-out-of-$N$ key partitions of equal size.**

The evaluation in [30] shows that the following RMI architecture outperforms B-Trees in speed and memory utilization: a *two-stage* RMI architecture with *tree structure* between models and a partition of non-overlapping keyset of equal size assigned to models on the leaves. For the root model, or first-stage model, the authors deploy a *neural network model* that can capture the coarse-grained shape of complex functions. The authors propose the use of a large number of leaves, i.e., fan-out in the order of thousands, each of which predicts the final memory location of the queried key. Each leaf represents a second-stage model that performs *linear regression* on the CDF of a subset of keys, i.e., a fast and storage-efficient model. For simplicity, we assume fixed-length records and logical paging over a continuous memory region.

**Prediction Error.** If the prediction is not accurate then a *local search* is called which discovers the record around the (falsely) predicted location. We emphasize here that if the prediction error is large, then the design needs to correct the error by performing an extended local search which translates into *accessing more memory location* which slows down the performance of the overall design.

## 3.2 Background on Poisoning Attacks

Let $D = \{x_i, y_i\}_{i=1}^n$ denote the data used by a learning model, where the feature vector is $x \in \mathbb{R}^d$ and the response variable is $y \in \mathbb{R}$. In the linear regression model, the output is computed via a linear function $f(x, w, b) = w^T x + b$ with parameters $w \in \mathbb{R}^d$ and $b \in \mathbb{R}$. The parameters $w, b$ are chosen so as to minimize the loss function $\mathcal{L}(D, w, b) = \frac{1}{n}\sum_{i=1}^n (f(x_i, w, b) - y_i)^2$ which is the mean squared error (MSE). We note that regression differs from classification since the output is a numerical value, as opposed to a class label. Poisoning attack is described as a bilevel optimization problem: given the target model as the first-level minimization of the error function, the attacker aims to find new data that maximize the outcome of the first-level minimization. Previous works focus on *gradient-based* poisoning attacks. Some works propose alternatives to analytically solving the bilevel optimization problem, e.g., *sampling-based approach* [26] and *generative method* [61].

Our single-point poisoning attack in Section 4.3 takes a different approach by *exploiting the structure of CDFs* and computes the location of the poisoning point that maximizes the minimum error in a single pass over the legitimate keys. The multiple-point poisoning attack and the attack on RMI are extensions of our optimal and efficient poisoning approach.

## 3.3 Threat Model

**Attacker's Goal.** One of the main benefits of LIS is that they are significantly faster [30] than traditional index structures. In this work we focus on adversaries that insert maliciously crafted inputs to the set of keys of the LIS so as to corrupt the model and degrade its overall performance. The reason for such adversarial behavior depends on the adversarial gains and application context, e.g., competitor that wants to deteriorate performance or a denial of service attack. All recent pointers show that deploying technologies that are susceptible to adversarial manipulations is a dangerous approach with hard to predict consequences, e.g., extracting faces from models [16], corrupting autopilot models in cars [1]. In our case, the attacker's objective is to generate maliciously chosen training keys, called *poisoning keys*, that together with the *legitimate keys* will train an LIS that has prediction accuracy *lower* compared to an LIS trained only on legitimate keys. Poisoning attacks can be categorized into *poisoning integrity attacks* and *poisoning availability attacks*. Integrity attacks form a loss function over specific datapoints of interest and, therefore, aim for a targeted mis-prediction on this data. Availability attacks, which is the focus of this work, aim to indiscriminately deteriorate the performance.

**Attacker's Knowledge.** We focus on *white-box* attacks where the attacker has access to the training data, i.e., keyset $K$, and the parameters of the model, i.e, the $w$ and $b$ values for linear regression models. White-box attacks are a standard setting in robustness/privacy analysis of machine learning models [3, 4, 26, 32, 50, 51, 60, 61]. Specifically, the majority of academic work in LIS use publicly available datasets, e.g., [15, 17, 28, 30, 43], which makes such a white-box attack plausible. In practice, given the training data, an attacker can derive the parameters of the linear regression models by building the index locally before the computation of the poisoning keys. The scenario in which the attacker does not have access to the data or the parameters of the model, is called a *black-box* attack and it is outside the scope of this work. We note here that even if the exact training data is unknown, an attacker typically has information about the distribution which allows the construction of a *substitute dataset*. After generating a substitute dataset, an attacker can take advantage of the transferability of poisoning, which is shown to work well [60], and mount a black-box attack.

**Attacker Capabilities.** We assume the attacker can insert $p$ maliciously crafted poisoning keys before the training of the model. Poisoning attacks have been shown to corrupt the model in different scenarios outside LIS, e.g., recommendation systems [22] and crowdsourcing systems [11]. We denote the set of $p$ poisoning keys as $P$ and the overall *poisoned* keyset as $K \cup P$, which contains the sum of legitimate and poisoning keys, i.e., $n + p$. We call *poisoning percentage* the term $100 \cdot (p/n)$. We follow the footsteps of previous works in poisoning [3, 26] and investigate the effect of poisoning for up to 20% poisoning percentage. Jumping ahead, in our experiments we demonstrate the performance for a range of poisoning percentages and we observe that our attack is effective on small poisoning percentages as well.

**Attack Evaluation Metric.** The evaluation of the LIS performance in the original work by Kraska *et al.* [30] is performed by measuring the lookup time (in nanoseconds). The final benchmark numbers in their work are the result of custom code, designed for small models, which removes all unnecessary overhead and instrumentation that Tensorflow introduces in larger models. Unfortunately, the result of their engineering efforts are not publicly available and, therefore, we can not directly measure the effect of poisoning in LIS with respect to the time performance. In our work we evaluate the performance of our attacks with two implementation-independent metrics: (1) we compute the *Ratio Loss*, which is the ratio between the mean square error (MSE) function of the poisoned dataset and the MSE of the non-poisoned dataset; (2) we compute the *Average Memory Offset*, which is the average memory difference between the model prediction and true position of the associated record (in the unit of the memory size of a key-value pair).

## 4 POISONING REGRESSION MODELS ON CDF

In the paradigm of learned index structures (LIS) [30] the location of a key-record pair is computed by *approximating the relative order*, i.e., the rank, of the queried key. An accurate approximation of the rank allows the algorithm to jump directly to the desired memory location of the linearly ordered key-record pairs without touching the rest of the data in the index. In this section we propose attacks for *poisoning the linear regression model on CDF*, a building block for RMI. The proposed attack inserts points in the index with the goal of increasing the approximation error of the regression and as a result *degrade the time performance of the overall design*.

**A New Flavor of Poisoning.** For an index with keyset $K$ of size $n$, each key $k \in K$ has a rank $r$ in the interval $[1, n]$, which corresponds to its position in the sorted sequence of $K$. LIS *approximates the rank* of a key by a linear regression model on the pair $(k, r)$, where the $X$-value is the key and the $Y$-value is its rank. The model approximates the non-normalized cumulative distribution function.

In the traditional poisoning attacks on regression models [26], i.e., not on CDF functions, the insertion of a poisoning point causes only a "local" change since it *does not affect* the $X$- and $Y$-values of any of the legitimate points. On the contrary, for the case of LIS, the insertion of a single key $k_p$ changes the rank of all the legitimate keys larger than $k_p$. Consequently, this "global" change on the rank of the legitimate keys triggers a change on the CDF itself. This *compound effect of an adversarial insertion* has not been analyzed before. Another difference is that the rank of a chosen poisoning key $k_p$ is automatically implied by the other keys of the data set. Thus, the attacker can not pick arbitrary $X$-, $Y$-value assignments which is a major departure from the setting of all previous poisoning attacks. In this work, we introduce a new flavor of poisoning attacks on learned index structures.

Before addressing the general LIS design, which is comprised of a two-stage architecture, we first pose a fundamental question that focuses on the new flavor of poisoning: What is the optimal poisoning strategy that maximizes the error of linear regression applied on a CDF? We answer this question by developing an efficient attack that maximizes the error of a linear regression on a CDF.

The rest of this section is organized as follows. Section 4.1 formalizes the problem statement. In Section 4.2, we provide an intuitive explanation of our attack strategy. A detailed description of our attack method and analysis of its performance and complexity are given in Sections 4.3–4.4. Finally, Section 4.5 contains an experimental evaluation of the practical effectiveness of our attack.

### 4.1 Problem Statement

We first define the linear regression framework on cumulative density functions. The following definition bridges the notions of ranks and their corresponding CDF.

> **Definition 1 (Linear Regression on CDFs).** *Let $K = \{k_1, \cdots, k_n\} \subseteq \mathcal{K}$ be the set of integers that correspond to the keys of the index. Every key $k_i \in K$ has its associated rank $r_i \in [1, n]$. The* linear regression model on a CDF *computes a pair of regression parameters $(w, b)$ that minimizes the following mean squared error (MSE) function :*
> 
> $$\min_{w,b} \mathcal{L}\left(\{k_i, r_i\}_{i=1}^n, w, b\right) = \min_{w,b} \left(\sum_{i=1}^n (wk_i + b - r_i)^2\right).$$

In this work we focus on *non-regularized linear regression*, much like the original work on LIS by Kraska *et al.* [30]. The goal of regularization is to generalize the model on unseen (testing) data; in LIS the majority of queries are expected to be data stored in the index structure, i.e., training data. Therefore, the impact of regularization is unclear in the context of LIS.

We can derive a closed-form solution to the minimization problem of Definition 1. Notice that the set $K$ can be interpreted as a sample from the set of keys $\mathcal{K}$. Given this probabilistic point of view, we define the *sample mean* of the key set as $M_K$ and the sample mean of the rank set as $M_R$. We define the *sample variance* as $\text{Var}_K$ and $\text{Var}_R$, and the *sample covariance* between $K$ and $R$ as $\text{Cov}_{KR}$. Lastly, the sample mean of the squares of the keys, resp. ranks, is defined as $M_{K^2}$, resp. $M_{R^2}$. Recall that the formulas of variance and covariance are, $\text{Cov}_{XY} = M_{XY} - M_X M_Y$, $\text{Var}_X = M_{X^2} - M_X^2$.

THEOREM 1. [42] *The Linear Regression from Definition 1 admits the following closed-form solution:*

$$w^* = \frac{\text{Cov}_{KR}}{\text{Var}_K}, \quad b^* = M_R - w^* M_K, \quad \mathcal{L}(K, R, w^*, b^*) = -\frac{\text{Cov}_{KR}^2}{\text{Var}_K} + \text{Var}_K.$$

The adversarial goal of the newly introduced poisoning for linear regression on CDF is described in the following.

> DEFINITION 2 (**POISONING LINEAR REGRESSION ON CDF**).
> *Let $K$ be the set of $n$ integers that correspond to the keys and let $P$ be the set of $p$ integers that comprise the poisoning points. The augmented set on which the linear regression model is trained is $\{(k'_1, r'_1), (k'_2, r'_2), \cdots, (k'_{n'}, r'_{n'})\}$, where $k'_i \in K \cup P$ and $r'_i \in [1, n+p]$. The goal of the adversary is to choose a set $P$ of size at most $\lambda$ so as to maximize the loss function of the augmented set $K \cup P$ which is equivalent to solving the bilevel optimization problem:*
> $$\arg\max_{P \text{ s.t. } |P| \leq \lambda} \left( \min_{w,b} \mathcal{L}\left(\{k'_i, r'_i\}_{i=1}^{n+p}, w, b\right) \right)$$

The upper bound $\lambda$ in the size of $P$ is chosen to be proportional to the size of the keyset, e.g., $\lambda = 0.2n$. We define the notions of *number sequence* and *discrete derivative* [40] as: a number sequence $A$ is an ordered list of numbers and we denote with $A(i)$ the $i$-th number in the sequence.

DEFINITION 3. [*Discrete Derivative*] [40] *A discrete derivative of a sequence $A$ is defined as the difference between consecutive numbers in the sequence $A$. We denote the sequence of discrete derivatives of $A$ as $\Delta A$. Formally:*

$$\Delta A(i) = A(i+1) - A(i)$$

## 4.2 The Compound Effect of Poisoning CDF

For the sake of illustration we use a naive brute force approach to poison a key set with $n = 10$ keys. Figure 2-(A) shows the original index key set on the $X$-axis and the corresponding ranks on the $Y$-axis, while Figure 2-(B) shows the regression line after the poisoning. The blue vertical lines indicate the *distance of the point from the regression line*, i.e., the error incurred by this key. Therefore, one can illustrate the MSE by summing the square of the above distances. The key $k_p$, colored in red, is the optimal poisoning location. Due to the compound effect of an insertion on the CDF, the ranks, i.e. the $Y$-value, of the points after $k_p$ increase by one.

In a typical setting of poisoning regression models, the addition of a single point has a limited overall impact since all the other points stay in their original $X$-, $Y$-coordinates. On the contrary, for the case of CDFs the addition of a single point can affect the rank, i.e., $Y$-coordinates, of many original points of the CDF. In other words, a single point might force the regression line to *incur a compound error* from a large portion of the original points. The example of Figure 2 demonstrates this phenomenon, notice that the error-contribution by the majority of the keys, depicted with vertical blue segments is significantly larger in subfigure (B).

We now illustrate why previous poisoning attacks fail in the CDF setting. If we were to feed this 10-key dataset to a poisoning algorithm that neglects the fact that $Y$-dimension describes a CDF, then the optimal poisoning point would be (40, 1). Notice though that the poisoning key $k = 40$ would be assigned the rank 11, and not 1, since it is the greatest key. Therefore, the outputs of previous poisoning attacks can not deal with the characteristics of a CDF.

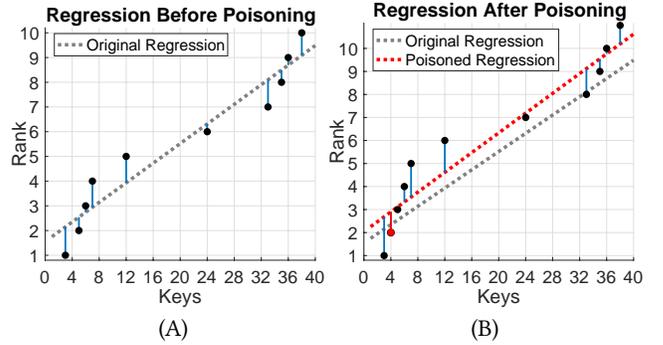

Figure 2: Illustration of the compound effect of poisoning using a single key $k_p$ colored in red. All original keys that are larger than $k_p$ increase their rank by one. The new regression line, dotted red line, accumulates larger error from most of the original points due to the adjustment of ranks.

## 4.3 Single-Point Poisoning on CDF

This subsection proposes an *efficient* attack for a single poisoning point that has strong evidence of being optimal. To avoid inserting out-of-range keys or introducing outliers (both poisoning approaches are detected and eliminated by known mitigations), we deem as potential poisoning keys the ones that lie in-between the smallest and the largest legitimate key.

*A First Attempt.* Since the key space is finite, we are guaranteed to find the optimal poisoning key if we compute the updated loss function for every potential poisoning key. For each poisoning, it takes $O(n)$ time to compute the loss from scratch, where $n$ is the number of existing keys. Since there are $m - n$ possible locations the time complexity of this approach is $O(mn)$. For large indexes with millions of keys and a large keyspace this approach is not practical.

*Our Approach.* The efficiency and optimality of our poisoning attack is based on the following observations about the problem:

**1) The loss function $\mathcal{L}$ can be seen as a sequence.** For a fixed keyset $K$, the loss function after poisoning boils down to a function that only depends on the location of the poisoning key $k_p$. Therefore, one can see the loss function $\mathcal{L}$ as a sequence denoted as $L$ where its index represents the location of the poisoning key $k_p$, and its output, denoted as $L(k_p)$, represents the MSE if we were to choose $k_p$ as the poisoning key. Since multiplicities are not allowed we get the following expression for the sequence:

$$L(k_p) = \begin{cases} \min_{w,b} \left( \sum_{k' \in K \cup k_p} (wk' + b - r')^2 \right) & , \text{ if } k_p \notin K \\ \perp & , \text{ if } k_p \in K \end{cases} \quad (1)$$

Likewise, the mean of keys $M_K(k_p)$ *is a sequence* where each value is the mean of the poisoned keyset with respect to the chosen poisoning key. The mean of the new ranks $M_R(k_p)$, the variance of the poisoned keys $\text{Var}_K(k_p)$, and the covariance between the poisoned keys and the new ranks $\text{Cov}_{KR}(k_p)$ are all sequences with the poisoning key as their index.

**2) The value of $L(k_p)$ can be re-used for $L(k_p+1)$.** In the "first attempt", one computes the new loss function on each potential poisoning key by processing the entire dataset, i.e., time complexity $O(n)$ for every evaluation. Our insight is that we can pay the linear cost once, i.e., $O(n)$ for the first potential poisoning key, and then *compute the new loss function for the next poisoning key in constant time*. To compute the loss for consecutive keys we use the notion of *discrete derivative* of the sequence $L(k_p)$, denoted as $\Delta L(k_p)$. Using the expressions for the optimal parameters from Theorem 1 we can directly compute the new loss function with optimal parameters for the *updated regression model*. The discrete gradients of the sequences are computed in constant time. Therefore, we can compute the value of the entire sequence for the loss function in $O(m+n)$ time, as opposed to $O(mn)$ of the "first attempt".

**3) Loss function is the composition of convex subsequences.** As identical keys are not allowed in the index, the domain of the entire loss sequence is comprised of *subsequences*. The domain of each subsequence consists of consecutive poisoning keys. The existing keys $K$ divide the key space into at most $(n-1)$ subsequences. The rank of a potential poisoning key is the same within each subsequence. We conjecture that each subsequence is convex with respect to the evaluation of the loss function.

CONJECTURE 1. *Let $K$ be the set of original keys. Let $L(k_p)$ be the sequence where for input $k_p$ it outputs the value of the loss function on a linear regression model trained on $K \cup k_p$. The loss sequence $L(k_p)$ is convex on the domain defined between each consecutive pair of key values $k_i$ and $k_{i+1}$ from $K$.*

Our conjecture is based on the following observation: the convexity on loss function $L(k_p)$ is a result of the convexity of $L'(k_p, w, b) = L(k_p)$, in which model parameter $w$ and $b$ are viewed as variables. The loss term is a summation of a series of $n$ quadratic functions on $w$ and $b$ with the exception of one term $(wk_p + b - r_p)^2$, where $wk_p$ is a nonlinear variable term. However, the value of this term is dominated by the summation of the other $n-1$ quadratic terms. Besides the theoretical intuition, in every experiment we performed, the loss function was convex. A corollary of the above conjecture is that, the maximum loss for each convex subsequence is given either by the first or the last poisoning key, i.e., the endpoints. Therefore, we can compute the global maximum of the entire loss sequence by a *constant-time computation for each subsequence* which reduces the time complexity of the computation from $O(m+n)$ to $O(n)$. Figure 3 presents the loss function across the key space; the convexity can be seen by the first derivative plot of this example.

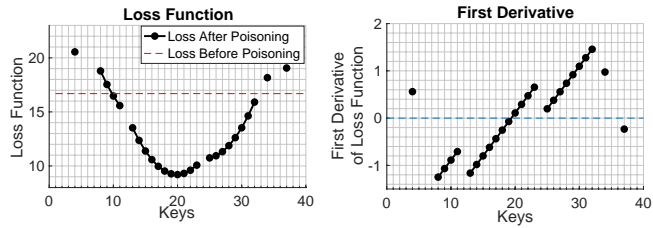

**Figure 3: An illustration of the loss function and its first derivative from the keyset of Figure 2. Each subsequence of consecutive poisoning keys is convex w.r.t. the loss function.**

**Single-point Poisoning.** We put the above observations together to define the closed-form formulas that compute the evaluation of the loss function for the poisoned set of keys. Let $(k_1, r_1), \ldots, (k_n, r_n)$ be the sequence of pairs of keys and ranks for a dataset. The set of keys $k_1, \ldots, k_n$ imply a collection of subsequences in the key domain such that each subsequence is comprised of *consecutive non-occupied keys*. For example, consider the key-rank pairs:

$$(k_1, r_1), (k_2, r_2), (k_3, r_3), (k_4, r_4) \leftarrow (2,1), (6,2), (7,3), (12,4).$$

The subsequences of non-occupied keys for the key domain $\mathcal{K} = [1,13]$ are: $\{1\}, \{3, 4, 5\}, \{8, 9, 10, 11\}, \{13\}$. Due to observation 3) we only need to consider the *endpoints of each subsequence*, i.e., for our example the endpoints are $\{1\}, \{3, 5\}, \{8, 11\}, \{13\}$. We define sequence $S$ where the element $S(i)$ corresponds to the $i$-th smallest key among *all the endpoints from all subsequences*. For our running example we have $S(1) = 1, S(2) = 3, S(3) = 5, S(4) = 8, S(5) = 11, S(6) = 13$. We also define the sequence $T$ where the element $T(i)$ corresponds to the rank that the poisoning key $S(i)$ would take if it is inserted. For our example we have $T(1) = 1, T(2) = 2, T(4) = 2, T(5) = 4, T(6) = 4, T(7) = 5$. To simplify the notation we index all the subsequences with respect to the index of sequence $S$, i.e., the notation $M_K(i)$ is equivalent to $M_K(k)$ where $k \leftarrow S(i)$.

The attacker first calculates the effect of inserting the first potential poisoning key $S(1)$, which implies the calculation of the values $M_K(1), M_{K^2}(1), M_{KR}(1)$, and $L(1)$. The effect of inserting poisoning key $S(i+1)$ on the loss function, i.e., $L(i+1)$, can be computed in constant time as:

$$M_K(i+1) = M_K(i) + \frac{\Delta S(i)}{n+1}, \quad M_{K^2}(i+1) = M_{K^2}(i) + \frac{(2S(i) + \Delta S(i))\Delta S(i)}{n+1}$$

$$M_R(i) = \frac{n+2}{2}, \quad M_{R^2}(i) = \frac{(n+2)(2n+3)}{6}, \quad M_{KR}(i+1) = M_{KR}(i) + \frac{T(i)\Delta S(i)}{(n+1)}$$

$$L(i+1) = -\frac{(M_{KR}(i+1) - M_K(i+1)M_R(i+1))^2}{M_{K^2}(i+1) - (M_K(i+1))^2} + M_{R^2}(i+1) - (M_R(i+1))^2$$

**Algorithm complexity.** The algorithm for the single-point poisoning runs as a subroutine in the Algorithm 1, see Lines 3-10. The above approach *maximizes the error of the poisoning for a single insertion* in $O(n)$ time complexity. Space complexity is constant for each single-point poisoning call. In each iteration, the algorithm accesses an item from the *endpoint sequence $S$* and updates the maximal loss. Since we only store the endpoint key value of each iteration, the memory usage is constant throughout the execution.

**Comparison with Previous Poisoning Attacks.** There are a few similarities but, more importantly, several differences between our approach and the state of the art result in poisoning for linear regression [26]. We elaborate in the following:

• *Similarities:* Both [26] and our work formulate the poisoning attack as a bilevel optimization problem. Additionally, both works deploy an iterative method to derive the poisoning points.

• *Differences:* The first difference concerns the *dimension* of the poisoning points. Specifically, when it comes to poisoning a CDF, the rank of each key depends on the values of all the keys of the index. Thus, to poison a CDF, the attacker chooses a one-dimensional point. Whereas in [26], the attacker chooses a two-dimensional point. We cannot apply the attack from [26] to our CDF setting because their algorithm (falsely) assumes that it can pick arbitrary ranking which creates inconsistent and unusable poisoning points, i.e., it chooses both $X$- and $Y$-coordinates. The above simple observation introduces a another technical difficulty that concerns the *gradient formula*. That is, the second difference is that the gradient of the optimization problem in [26] simply requires to plug in the poisoning $(x, y)$ coordinates both of which are chosen by the adversary. On the contrary, in the CDF setting, in order to compute the gradient for an arbitrary point the adversary has to access a (potentially) large number of keys to calculate the $Y$-coordinate since it corresponds to the rank of the new poisoning point. Thus, a naive poisoning for CDFs has to pay linear time for every gradient computation which is why we devised a tailored amortized analysis for poisoning in the CDF setting in Section 4.3. We achieve a "constant-time" computation because we iterate through potential poisoning points with +/-1 change in ranking, our efficient method cannot be applied to arbitrary steps of an iterative algorithm such as [26]. Finally, as discussed in Conjecture 1, the optimization problem of poisoning a CDF has a *global* maximum, whereas the attack of [26] may get trapped in a local optimum.

---

**Algorithm 1:** GreedyPoisoningRegressionCDF

**Data:** The number of allowed poisoning keys $p$, the original dataset for the regression $\{(k_1, r_1), \ldots, (k_n, r_n)\}$ where $k_i \in K$ and $r_i \in [1, n]$.
**Result:** Set of poisoning keys $P$ such that $P \cap K = \emptyset$ and $|P| = p$.

1 Initialize the set of poisoning keys $P \leftarrow \emptyset$;
   // Follow a greedy approach, choose locally optimal poisoning
2 **for** *every $j$ from 1 to $p$* **do**
3    Partition the non-occupied keys, i.e., keys not in $K \cup P$, into subsequences such that each subsequence consists of *consecutive* non-occupied keys;
   // Due to convexity, the loss function is maximized at an endpoint
4    Extract the endpoints of each subsequence and sort them to construct the new sequence of endpoints $S(i)$, where $i \leq 2(n + j)$;
5    Compute the rank that key $S(i)$ would have if it was inserted in $K \cup P$ and assign this rank as the $i$-th element of the new sequence $T(i)$, where $i \leq 2(n + j)$;
   // Evaluate each sequence for the smallest endpoint
6    Compute the effect of choosing $S(1)$ as a poisoning key and inserting it to $K \cup P$ with the appropriate rank adjustments. Specifically, evaluate the sequences each of which is the mean $M$ for a different variable, e.g., $K, R, KR$. Compute $M_K(1), M_{K^2}(1), M_{KR}(1)$, and $L(1)$ ;
7    **for** *every $i$ from 2 to the length of sequence $S$* **do**
8       Compute the effect of choosing $S(i + 1)$ as a poisoning key by calculating the loss function $L(i + 1)$ from the equations in (4.3);
9    **end**
10    Define as $k_{\text{OPT}} \leftarrow S(\arg\max_i L(i))$ the chosen poisoning key which maximizes the loss;
11    Augment $P$ as $P \leftarrow P \cup k_{\text{OPT}}$ ;
12 **end**
13 **return** *the set of poisoning keys $P$*;

---

### 4.4 Greedy Multiple-Point Poisoning on CDF

We generalize the single-point approach so as to insert multiple poisoning keys. Specifically, we propose a greedy approach where at each iteration the attacker makes a locally optimal decision and inserts the poisoning key that maximizes the error of the augmented keyset so far, see Algorithm 1. Even though we do not provide a proof of optimality for the multiple-point poisoning, we experimentally observed that our approach matched the performance of the brute-force attack in every tested dataset. Intuitively, our approach places poisoning keys in a dense area so as to *exacerbate the non-linearity of the CDF* and consequently increase the error.

### 4.5 Evaluation

In this subsection we evaluate the effect of greedy multi-point poisoning. We observe that the ratio loss increases up to 100× depending on the size of the dataset and the domain of the keys. Besides the loss function, we also observed that the poisoned index needs to access significantly more memory locations, i.e., via the correctness process of the so-called local search of LIS. Specifically, we observed that after poisoning, every query has to access (on average) 5% of the entire dataset to correct the accuracy error. That is 10× larger number of accesses than the non-poisoned dataset.

**Setup.** We produce synthetic datasets of keys that are uniformly distributed. We note here that the uniform distribution has small MSE loss because of the near-linear CDF due to uniformity. This implies that these datasets are the ones that the linear regression on CDF can capture well and, thus, one of the cases where learned index structures outperform traditional methods. We chose different parameters for our experiments. The first is the *number of legitimate keys* (denoted as "Keys"), the second is the *density* of the legitimate keys over the key domain. We evaluate on the following fixed densities: 5%, 10%, 50% and 80%. The third, which can be computed from the first two, is the *size of the key domain*, and the fourth is the *percentage of poisoning keys* with respect to the number of legitimate keys. Following the footsteps of previous poisoning attacks [26] we only consider poisoning percentage up to 15%. We note that in our experiments we fix the number of keys and the density and *adjust the key domain* accordingly. The reason behind this design choice has to do with the architecture of the original work on LIS [30]. Specifically, in the two-level architecture of RMI the model partitions the entire keyspace so that each partition has *a fixed number of keys* and as a result the size of the key domain varies between partitions, as opposed to a fixed key domain size with varying number of keys. The chosen key domain sizes in our experiments were picked so as to follow typical domain sizes where the regression on LIS was performed in the original work [30]. To measure the effectiveness of our multi-point poisoning attack, we record *Ratio Loss*, which is the ratio of the MSE loss on the union of poisoning and legitimate keys, over the MSE loss on just legitimate keys. We also record *Average Memory Offset*, which is the average offset between the predicted location and real location of the key. E.g., memory offset 100 means that the local search has to go through 100 key-value pairs to correct the error.

**Results.** The results of our evaluation are presented in Figure 4. For a fixed number of legitimate keys and a fixed key density, i.e., focusing on a single plot of Figure 4 (a), we see that the higher the poisoning percentage the larger the ratio loss and the average memory offset. For instance, in the large key domains we see that the ratio increases up to 100× as the poisoning percentage gets larger. On the other hand, we also observe that when the density is too high, it may result in a much smaller error increase in both measurements. Another interesting observation from our experiments is that lower density for the same fixed number of legitimate keys implies a larger set of potential poisoning keys and, thus, allows for a greater increase of error. This can be seen from the drop of the ratio loss on a fixed row of the Figure 4 (a). The above intuition is confirmed even when we compare a column of Figure 4 (a), i.e., fix the key density and increase the domain. In that case we observe that larger domains with the same fixed density also imply a larger set of potential poisoning keys and thus allows for a greater increase of error. We also observe that the average memory offset, Figure 4 (b), grows linearly with the size of dataset. For example, in the last boxplot of the bottom left plot, the linear regression on average predicts a memory address that is 200× the size of a single key-value pair afar from the true address.

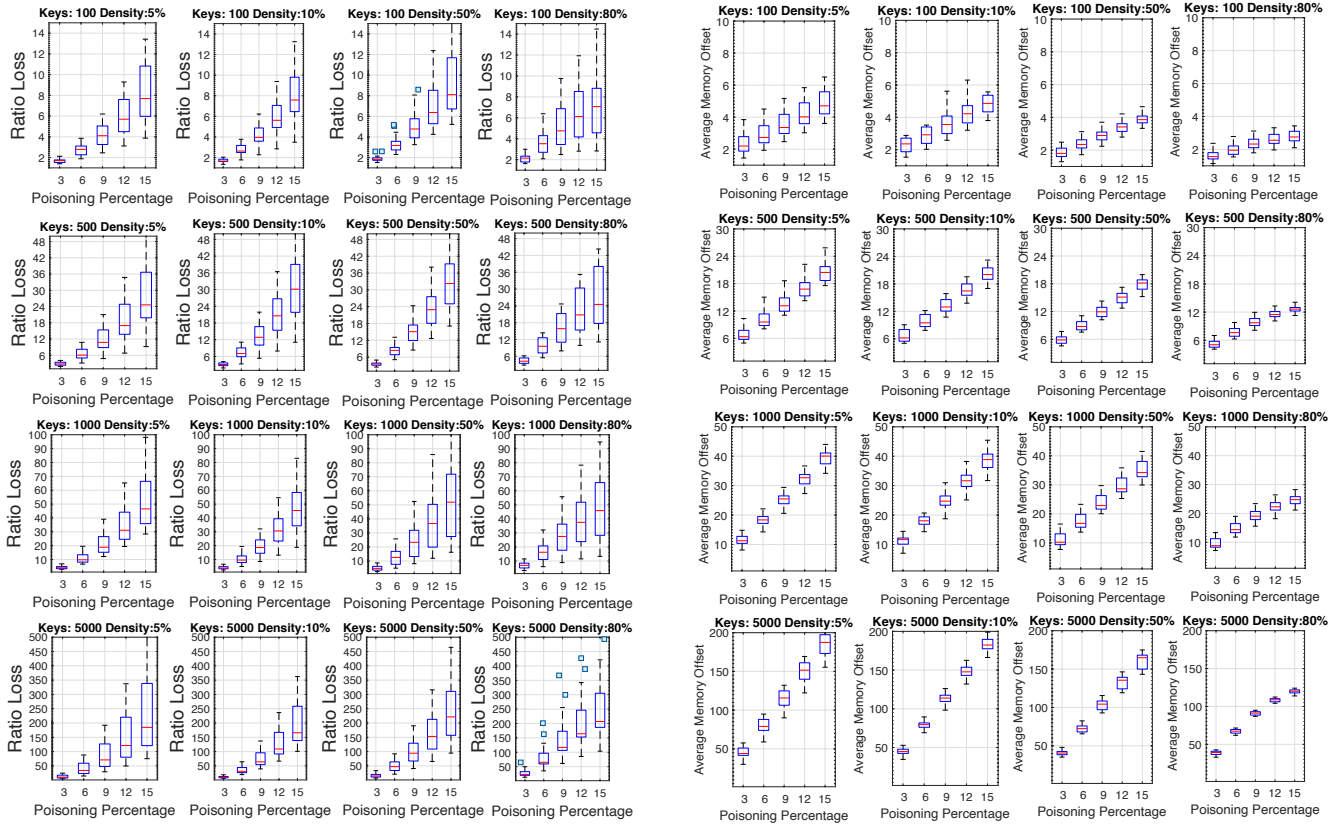

Figure 4: Evaluation of the multi-point poisoning for linear regression on CDF. The first set evaluates the *Ratio Loss,* and the second set evaluates the *Average Memory Offset*. Each boxplot shows the respective loss on the legitimate keyset over 20 distinct keysets. The legitimate keys are distributed uniformly. The number of legitimate keys, denoted as "Keys" and the size of the key domain, denoted as "Key Domain" are presented on the title of each plot. The number of poisoning keys varies on the $X$-axis and is represented as a poisoning percentage with respect to the number of legitimate keys.

**Why Uniformly Distributed Data?** The superior performance of RMI relies (in part) on the fact that it partitions the keyset in small consecutive keysets that are roughly similarly distributed. Therefore, even if the overall CDF has structure that is hard to capture, each "local" structure is similar enough to be captured with a simple model. Based on the above rationale we run experiments on a uniformly distributed keyset in order to capture a typical local model of RMI. For completeness, we also performed experiments on a dataset generated with a normal distribution; we report our findings in the Appendix of this work. For a key domain $U = [\alpha, \beta]$ we form the normal distribution with mean $\mu = \frac{\beta+\alpha}{2}$ and standard deviation $\sigma = \frac{\beta-\alpha}{3}$. Because the normal distributions are not captured well by linear models, the MSE loss of the original dataset is high. Despite that, our attack achieves up to 8× increase of error.

**On Local Search Strategies.** Kraska *et al.* propose two local search strategies: linear search and binary search. In case of linear search, the average memory offset corresponds to the amount of memory that needs to be scanned to correct the accuracy error. In case of a binary search, LIS take the maximum memory offset from the entire dataset in order to "bound" the search area they perform the binary search algorithm on. For completeness we provide additional experiments that show the *maximum memory offset* to address the case of a binary search based local search. Instead of measuring the average memory offset for each dataset, we measure the maximum memory offset over all queries on the keys that present in the dataset. Similar to average memory offset, maximum memory offset grows linearly with the size of the dataset. The maximum memory offset is roughly two times the average memory offset given the same setup, which implies that the skew of the distribution of memory offsets is close to center.

# 5 POISONING TWO-STAGE RMI MODELS

Armed with the poisoning technique for regression on CDF from Section 4, in this section, we develop a poisoning attack for the two-stage hierarchical model of recursive model index (RMI). Our approach is tailored to the index architecture proposed by Kraska *et al.* [30]. We evaluate our attack on RMI on synthetic and real data under various scenarios.

**Structure of Second-Stage Models.** According to Kraska *et al.* [30], an index architecture that outperforms traditional B-Trees is comprised of a neural network for the first-stage model and a linear regression on a CDF for the second-stage model. As described in Section 3.1, each second-stage model is the "expert" in fine-tuning the prediction on a *fixed subset of keys*. During the initialization of the RMI, the designer partitions the set of all keys, $K$, into $N$ non-overlapping subsets $K_1, \ldots, K_N$ of equal size. With $R_i$ we denote the set of ranks for the corresponding subset of keys $K_i$ of the $i$-th second-stage model. As a next step the designer independently trains a linear regression on each $\{K_i, R_i\}$ and stores the regression parameters $w_i, b_i$. Jumping ahead, our attack exploits the fact that a key-partition step takes place to decide which second-stage models to poison and how many keys to inject.

**Loss Function for RMI.** In this work, we do not consider the case of poisoning the neural network (NN) model of the first-stage. The reason behind this design choice is that according to the experiments in [30], a query key that is used in the training of the NN model will always direct to the correct regression model. Therefore, we focus our poisoning attack on manipulating the models of the second-stage and assume that the NN model will always point to the correct (albeit poisoned) second-stage model. Let $\mathcal{L}(\{K_i, R_i\}, w_i, b_i)$ denote the loss function of the $i$-th linear regression model of the second-stage that is trained on the CDF of $\{K_i, R_i\}$ and has parameters $w_i, b_i$. We define the loss function of the RMI as:

$$\mathcal{L}_{\text{RMI}}(K) = \frac{1}{N} \sum_{i=1}^{N} \mathcal{L}(\{K_i, R_i\}, w_i, b_i).$$

**Poisoning Threshold per Regression Model.** An important observation is that the adversary controls (1) which regression models to poison among the second-stage models, and (2) how many poisoning keys to inject in each regression model. We argue that injecting too many poisoning keys in a single regression model might allow a defense mechanism to detect such a behavior. In our attack, we handle this issue by imposing a poisoning threshold for each individual regression model, denoted by $t$. Recall that the term poisoning percentage, see Section 3.3, controls the *total*, as opposed to per model, number of poisoning keys.

A first attempt on this attack scenario is to pick a fixed poisoning threshold across all linear regression models to follow the overall poisoning percentage, $\phi$, i.e., $t = \frac{\phi n}{N}$. Such an approach allows only a single way of allocating poisoning keys to each model, e.g., if the poisoning percentage $\phi$ is 10% on a keyset of size $10^6$ with key partitions of size $10^3$, then the above approach can only assign 100 poisoning keys on each regression model. Thus, it is not possible to skew the assignment of poisoning keys. To increase the impact of the attack, we allow the poisoning threshold to vary across models provide it *does not exceed a certain upper bound*. In particular, denoting with $t_i$ the poisoning percentage of the $i$-the model, we require $t_i \leq t = \alpha \cdot \frac{\phi n}{N}$, where $\alpha$ is a small constant. For example, in our experiments we pick $\alpha \in \{2, 3\}$ which means that for 10% poisoning percentage in our previous example we allow *up to $t = 200$* (resp. $t = 300$) poisoning keys per regression. This approach allows multiple ways of assigning poisoning keys to regression models and, thus, allows our attack to achieve larger $\mathcal{L}_{\text{RMI}}$ error without overpopulating with poisoning keys the regression.

**Formulation for Poisoning RMI.** Let $P_i$ be the set of poisoning keys injected to the $i$-th model of the second-stage. Let $\phi$ be the allowed overall poisoning percentage for the RMI model and let $t$ be the poisoning threshold per second-stage model. Then, the goal of the poisoning attack on two stage RMI can be expressed as:

$$\arg\max_{P_1, \ldots, P_N} \sum_{i=1}^{N} \min_{w_i, b_i} \mathcal{L}(\{K_i \cup P_i, [1, n + |P_i|]\}, w_i, b_i)$$

such that, $|P_i| \leq t, \forall i \in [1, N]$, and $\sum_{i=1}^{N} |P_i| \leq \phi n$

We can re-frame the optimization into two subproblems. The first is to choose how many poisoning keys are injected per partition, we call this *volume allocation problem*, and the second is to choose which poisoning keys to inject within a specific partition, we call this the *key allocation problem*. For the latter problem we use Algorithm 1 for greedy poisoning the regression. The volume allocation problem is an $N$-dimensional integer programming problem with input vector $(|P_1|, \ldots, |P_N|)$. The search space is comprised of the volume assignments such that $|P_1| + \cdots + |P_N| = \phi n$ and the volume on each dimension is bounded by 0 and $t$. For realistic datasets it is infeasible to explore this search space. We propose a greedy approach for the volume allocation that performs well in practice.

## 5.1 Poisoning Algorithm

Given the discussed formulation for poisoning RMI we propose an attack that (A) follows a greedy approach for the volume allocation problem, and (B) given a fixed volume per second-stage model, it applies Algorithm 1 for the key allocation problem.

**Our Second-Stage Volume Allocation Approach.** In Algorithm 2, we follow a greedy approach that takes locally optimal steps for deciding how many poisoning keys to allocate, i.e., volume allocation, at each second-stage regression model. Note that our attacker has a total of $\phi n$ poisoning keys to allocate. As a first step the attacker distributes poisoning keys uniformly among models, i.e., $\phi n/N$ poisoning keys for each of the $N$ second-stage models.

The intuition for our greedy approach is that we *exchange a poisoning key for a legitimate key* between one regression model and its neighbor, i.e., either the next model or the previous model, if this re-allocation causes the maximum increase in loss function $\mathcal{L}_{\text{RMI}}$ among all the key-exchanges of this type.

We accompany the re-allocation of a poisoning key from the $i$-th model to the $j$-th model with the (reverse) move of a legitimate key from the $j$-th model to the $i$-th. The above step guarantees that the number of keys, i.e., the sum of poisoned and legitimate, stays fixed throughout the re-allocation moves. More formally we use the notation $i \rightarrow i + 1$ to indicate the exchange of keys where the direction of the arrow indicates the move of a poisoning key. Specifically, the first move is that a poisoning key that was available for placement in the $i$-th model now is allowed to be placed into the $(i + 1)$-th model instead, and the second move is that the *minimum* *legitimate key* of the $(i+1)$-th model is assigned to the $i$-th model. Similarly, the notation $i \leftarrow i+1$ indicates that a poisoning key that was available to the $(i+1)$-th model now moves to the $i$-th, and that the *maximum legitimate key* of the $i$-th model is assigned to the $(i+1)$-th model. For the purpose of Algorithm 2 we need to keep track of which reallocation of a poisoning key causes the maximum increase in $\mathcal{L}_{RMI}$. We define a simple two dimensional array denoted as ChangeLoss where entry ChangeLoss$(i, i+1)$ contains the change in loss $\mathcal{L}_{RMI}$ if the attacker executes the moves implied by $i \rightarrow i+1$ given the current state of allocation. Similarly, the notation ChangeLoss$(i+1, i)$ captures the reallocation $i \leftarrow i+1$. In every iteration, Algorithm 2 finds the maximum entry of ChangeLoss and applies the exchange, see Line 6, with the caveat that the addition of a poisoning key to $j$ does not violate the upper bound threshold $t$ of poisoning keys. Algorithm 2 terminates when the increase in $\mathcal{L}_{RMI}$ loss is less than $\epsilon$, see Line 5.

Performing the set of moves for exchange $i \rightarrow i+1$ (or $i \leftarrow i+1$) in Line 7, implies that a constant number of entries from ChangeLoss are rendered inconsistent since they rely on an old state of the volume allocation. Interestingly, because the exchange in keys takes place between models $i$ and $i+1$ it only affects the ChangeLoss entries of their direct neighbors, all other models stay unaffected with respect to their ChangeLoss entries. Without loss of generality, we assume that $i \rightarrow i+1$ was chosen. Then, since a new poisoning key is injected in the $(i+1)$-th model and a legitimate is removed we have to update all the entries that refer to $(i+1)$. Thus, we have to recompute the following entries: ChangeLoss$(i, i+1)$, ChangeLoss$(i+1, i)$, ChangeLoss$(i+1, i+2)$, ChangeLoss$(i+2, i+1)$. Additionally, we have to recompute the following entries of the $i$-th model that now has one less poisoning key and one more legitimate key: ChangeLoss$(i, i-1)$, ChangeLoss$(i-1, i)$. To compute the updated entry for ChangeLoss we run Algorithm 1 which takes time linear to the number of keys of the second-stage regression model, i.e., $O(n/N)$. Since there are six updates that take place for the chosen exchange $i \rightarrow i+1$, the complexity of Line 8 is $O(n/N)$. The space complexity is dominated by the maintenance of the array ChangeLoss that contains $2N$ elements.

## 5.2 Evaluation on Synthetic Data

Our results show that the RMI error after poisoning increases up to $150\times$ while the individual regression error increases up to $1000\times$.

**Setup.** In the experiments shown in Figure 5, we produce synthetic data sets of keys that are uniformly distributed in key domains $|\mathcal{K}| = 10^6$ and $|\mathcal{K}| = 10^8$. We implemented three different architecture for assigning legitimate keys to second-stage models, i.e., RMI architectures, so as to validate our approach across different efficiency and accuracy trade-offs for RMIs. The first scenario analyzes the case of a large number of second-stage models, specifically $10^5$ indicated with "#Models" on the title of the boxplot, where each is responsible for a small number of keys, specifically $10^2$ indicated with "Model Size". The above RMI architecture corresponds to the plots on the first column of Figure 5. In the second and third architectures we decrease the number of second-stage models which implies an increase in the number of keys per model. These architectures correspond to the boxplots on the second and third column of Figure 5. Overall, as we iterate through columns of Figure 5

---

**Algorithm 2:** GreedyPoisoningRMI

**Data:** Poisoning percentage $\phi$, number of second-stage models $N$, keyset $K = \{k_1, \ldots, k_n\}$ where $k_i \in \mathcal{K}$, termination bound $\epsilon$, poisoning threshold per regression model $t$

**Result:** Set of poisoning keys $P_1, \ldots, P_N$ such that $P_1 \cap \ldots \cap P_N \cap K = \emptyset$, $\sum_{i=1}^{N} |P_i| = \phi n$, and $|P_i| < t$.

// Initial Volume Allocation

1 Iterate through all the regression models of the second-stage and for the $i$-th model, initialize $P_i$ by injecting $\phi n/N$ poisoning keys using Algorithm 1;

2 Compute $\mathcal{L}_{RMI}$ by averaging the loss of second-stage models;

// Store effect of exchange $i \rightarrow i+1$ in ChangeLoss

3 Iterate through all second-stage models; for each model $i \in [1, N]$, compute the change in $\mathcal{L}_{RMI}$ if we were to (A) add a poisoning key to $P_{i+1}$, (B) move the smallest legitimate key from $(i+1)$-th model to the $i$-th, and (C) remove a poisoning key from $P_i$. Store the difference between the new $\mathcal{L}_{RMI}$ after the above moves and the current loss $\mathcal{L}_{RMI}$ in the entry ChangeLoss$(i, i+1)$;

// Store effect of exchange $i \leftarrow i+1$ in ChangeLoss

4 Iterate through all second-stage models; for each model $i \in [1, N]$, compute the change in $\mathcal{L}_{RMI}$ if we were to (A) add a poisoning key to $P_i$, (B) move the largest legitimate key from $i$-th model to the $(i+1)$-th, and (C) remove a poisoning key from $P_{i+1}$. Store the difference between the new $\mathcal{L}_{RMI}$ after the above moves and the current loss $\mathcal{L}_{RMI}$ in the entry ChangeLoss$(i+1, i)$;

// Greedy iteration that increase the loss of the RMI

5 **while** *the change in $\mathcal{L}_{RMI}$ is larger than $\epsilon$* **do**
   // Perform a Greedy exchange of a poisoning key with a legitimate key between consecutive models
6   Find the indices $i, j$ that (A) correspond to the largest entry ChangeLoss$(i, j)$ and (B) do not violate the poisoning threshold $t$ for the $j$-th model;
7   Perform the exchange $i \rightarrow j$ between models $i$ and $j$ and use Algorithm 1 for adding a poisoning key to $P_j$;
   // Fix the consistency of ChangeLoss as a constant number of entries from ChangeLoss were modified with respect to the previous volume allocation
8   Recompute the inconsistent entries of ChangeLoss, i.e., entries that address refer to $i$, or $j$, or both;
9 **end**
10 **return** *the set of poisoning keys* $P_1 \cup \ldots \cup P_N$;

---

from left-to-right, the RMI model decreases its storage overhead as well as its prediction accuracy of the second-stage model. Finally, we tested two different multiplicative constants $\alpha$ of poisoning thresholds for each individual second-stage model.

**Results.** The results of our evaluation are presented in Figure 5. Each boxplot shows the ratio loss across all second-stage linear regression models. We present the ratio loss for each regression model individually so as to provide a *more fine-grained analysis* of how the attack performs. We also present the ratio between the loss of the poisoned RMI model and the non-poisoned RMI with a black horizontal line. As is expected when the poisoning percentage increases the effectiveness of the attack increases as well. For a fixed row of boxplots in Figure 5, we see that the larger the second-stage model the better our attack performs. This phenomenon is explained by the fact that the linear regression is responsible for more data; therefore, there are more opportunities for the poisoning to increase its effectiveness. Given the same key distribution and setup we see that the size of the key domain did not affect significantly the performance of our poisoning attack on RMI, i.e., the loss for RMI is slightly larger for key domain $10^9$ compared to $5 \cdot 10^7$. Additionally, the difference between the error for poisoning threshold per regression $\alpha = 2$ and $\alpha = 3$ is not significant.

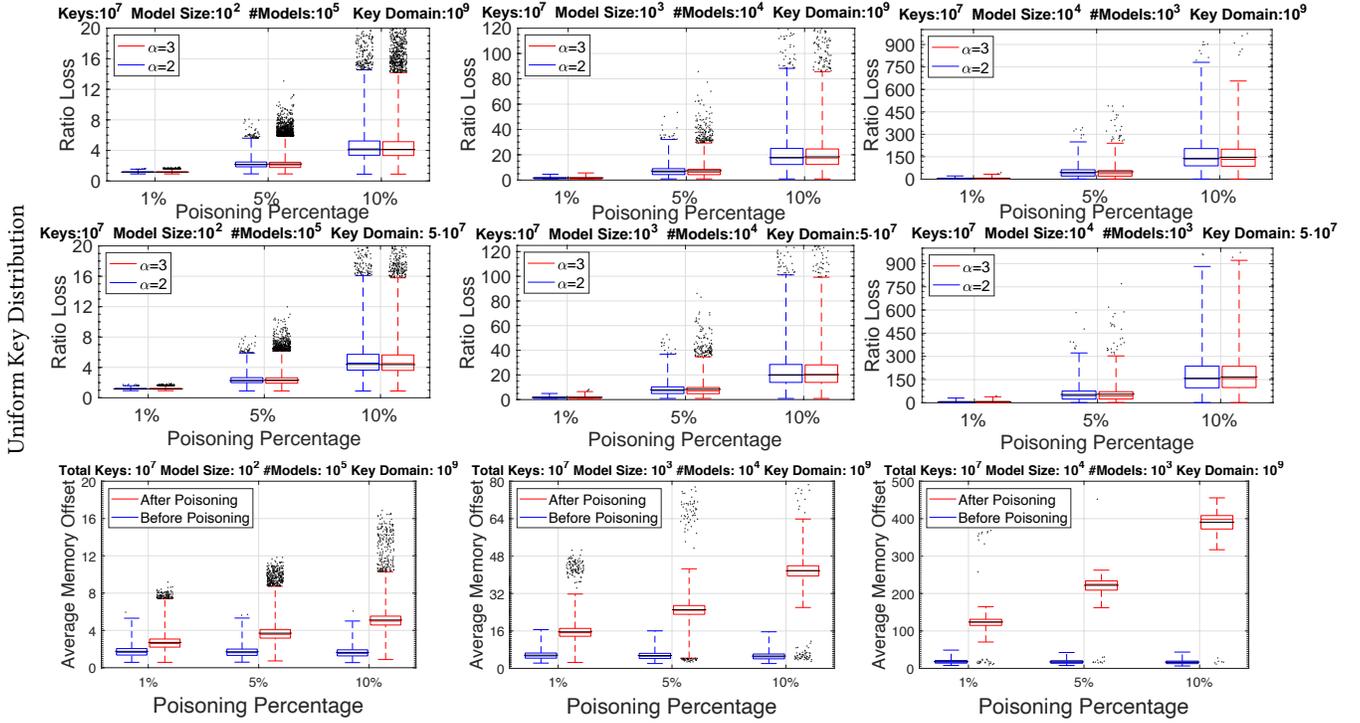

Figure 5: Evaluation of the multi-point poisoning for RMI on uniform distributed key set. The first two rows show the ratio loss and the third row shows the average memory offset. In the first two rows, the color of the boxplot denotes different $\alpha$ values for the poisoning threshold per model. In the last row, the blue boxes show the non-poisoned datasets and the red boxes show the corresponding datasets after poisoning. In all boxplots, the black line shows the respective error over the entire dataset.

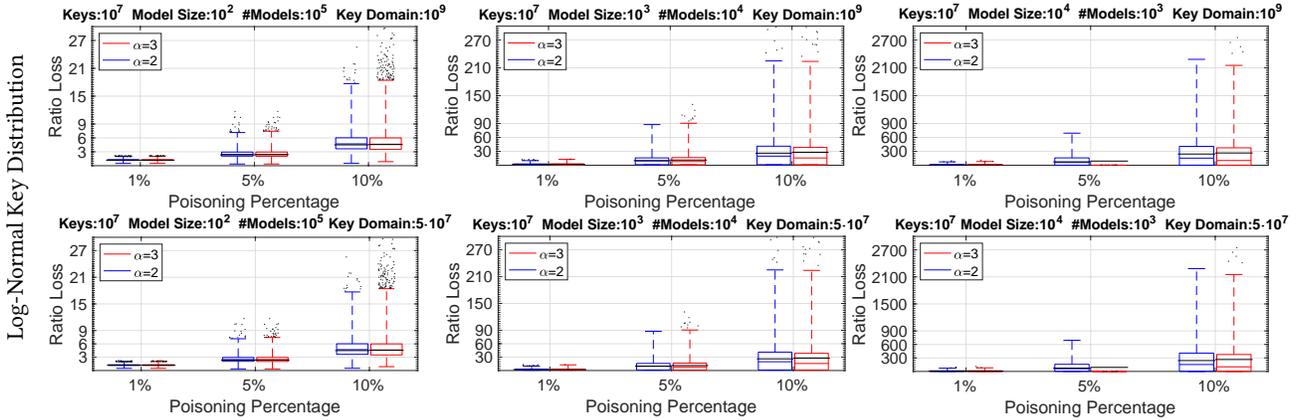

Figure 6: Evaluation of the multi-point poisoning for RMI on log-normal distributed key set. Each boxplot shows the ratio loss across all second-stage models. The ratio between the loss of the poisoned RMI model and the non-poisoned is represented with a black horizontal line. The $X$-axis represents different poisoning percentages and the color of the boxplot denotes different $\alpha$ values for the poisoning threshold per regression model.

**Additional Experiments.** Additionally, we tested our method on synthetic data sets of keys that are distributed with a log-normal distribution with $\mu = 0$ and $\sigma = 2$ (the same parameterization as the experiments in [30]). We illustrate the result in Figure 6. We observe that the performance of the attack is superior in the log-normal distribution compared to the uniform. In fact, the ratio loss is up to 2× larger for the same RMI setup. Interestingly, the whiskers of the boxplot are close to 3× larger for the case of the log-normal distribution, which implies a much larger spread of ratio loss values among the second-stage models. This is explained by the fact that in the log-normal case, we have some regressions that handle concentrated keys and by poisoning these models, we amplify the non-linearity of the legitimate keys which results in larger errors. In general, for the log-normal distribution, the RMI error presents up to 300× increase; whereas, in the individual second-stage level, we observed up to 3000× error increase.

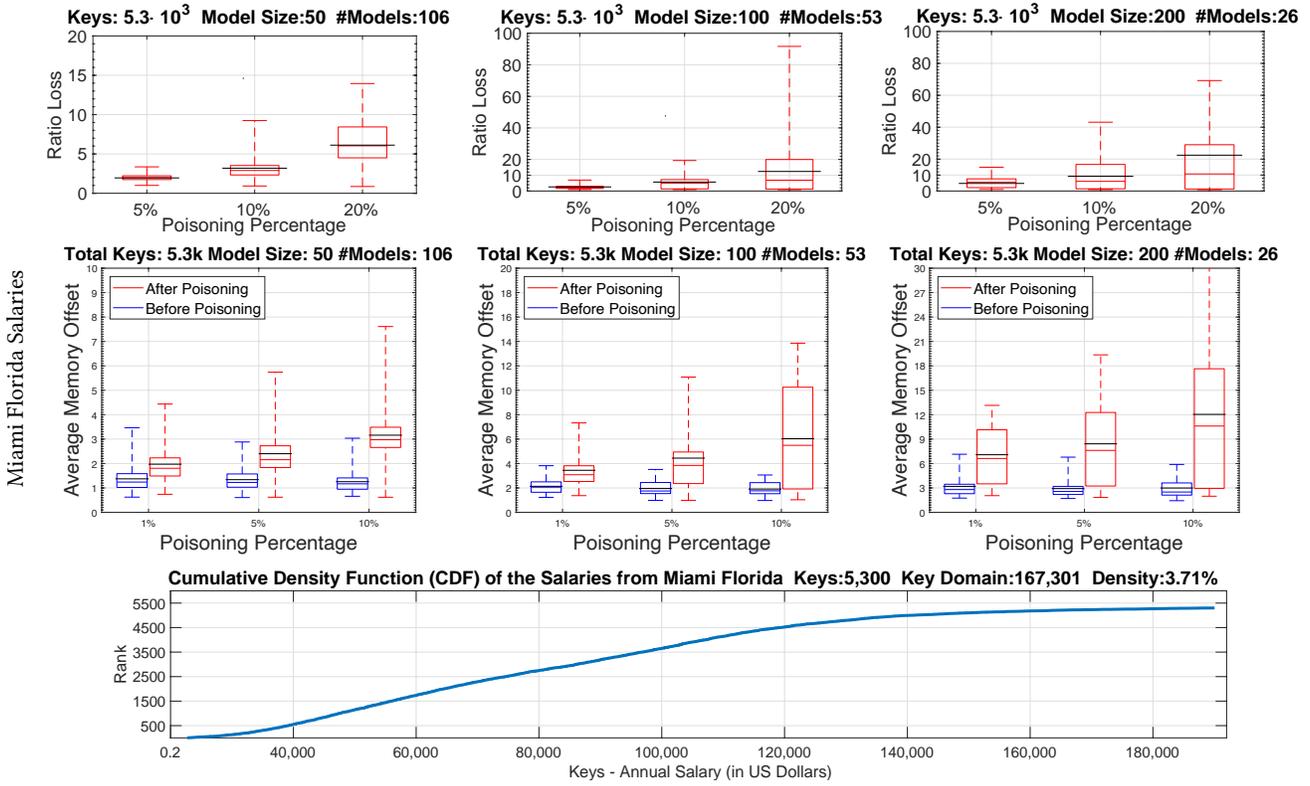

Figure 7: Evaluation of the multi-point poisoning for RMI applied on the CDF of the unique salaries of employees from Dada County in Miami. The $X$-axis represents different overall poisoning percentage where the second-stage poisoning threshold $\alpha$ takes value $\alpha = 3$. The third row presents the CDF.

## 5.3 Evaluation on Real-World Data

We evaluate the proposed greedy poisoning attack on RMI models on a real-world dataset of salaries of the employees of Dade County in Florida [38]. Our experiment show that the RMI error of a poisoned key set increases up to 24× and the error of the individual regression increases up to 70×.

**Setup.** We use the publicly available dataset of salaries of employees of Miami Dade County in Florida [38]. We only take unique salaries between $22,733 and $190,034, i.e., we pre-process to filter the outliers. The final dataset has $n = 5,300$ keys in a key universe of size $m = 167,301$ which gives a 3.71% key density. The CDF is depicted in Figure 7. In this experiment, we test three different RMI setups. In the first setup, we initialize the second-stage models with 50 keys each, which results in 106 second-stage models. In the second setup, we initialize the second-stage with 100 keys each, which results in 53 models. In the third setup, we initialize the second-stage with 200 keys each, which results in 26 models. For all the setups, we consider the poisoning threshold of the second stage model to have parameter $\alpha = 3$. For each experiment, we record both the ratio loss and the average memory offset. We consider poisoning percentages 5%, 10%, and 20%.

**Results.** The results of our evaluation are presented in Figure 7. For a fixed setup, we observe that as the poisoning percentage increases, the ratio loss increases as well. Across all setups, the ratio loss increases between 4× to 24×. The average memory offset of both the original dataset and the poisoned dataset scales with the size of second stage model. On all three setups, 10% poisoning causes the average memory offset to increase by 3×. Since the number of poisoning keys per second-stage model is a percentage over its number of keys, we see that larger models allow more poisoning and, consequently, larger RMI error. The discrepancy between the effectiveness of the attack on synthetic data and the real data is partly explained by the fact that the synthetic dataset is four orders of magnitude larger than the real dataset.

**Additional Experiments.** In the Appendix we present in detail experiments on the geolocation dataset [44] which is index by the latitudes of locations. We pick the data points that are labelled as schools over the world with latitudes between -30 and +50; and scale up the latitudes by 15,000 before rounding to achieve uniqueness of keys. The final dataset has $n = 302,973$ keys in a key universe of $m = 1,200,000$, which yields a 25% density.

In this experiment, we test three different RMI model setups. In the first setup, we initialize the second-stage models with 50 keys each, which translates to 6,059 second-stage models. In the second setup, we initialize the second-stage with 100 keys each, which translates to 3,029 models. In the third setup, we initialize the second-stage with 200 keys each, which translates to 1,514 models. For all the setups, we consider the poisoning threshold of the second stage model to have parameter $\alpha = 3$. Finally, we considered poisoning percentages 5%, 10%, and 20%.

For a fixed setup, we observe that as the poisoning percentage increases, the ratio loss increases as well. Among all experiments in both datasets, the RMI loss increases between 4× to 24×. Since the number of poisoning keys per second-stage model is a percentage over its number of keys, we see that larger models allow more poisoning and, consequently, larger RMI error.

## 6 DISCUSSION

**On Generalizing the Attack.** We present our attack on the design that initiated the study of learned systems [30]. Although our attack does not always transfers as is to all proposed LIS designs, we believe poisoning attacks mounted in a similar fashion are possible in most cases until we formally analyze the robustness of these models. The underlying theme of all learned index proposed so far is that a *model adapts to the underlying data*, e.g., via linear regression [7, 30] or via piece-wise linear approximation [15] or via linear interpolation [17] or via polynomial interpolation [47]. Given that a single maliciously chosen datapoint changes (i) the parameterization of the model, i.e., an ML model needs to retrain to take into account the error introduced by the addition, as well as (ii) the coordinates of up to a linear number of legitimate CDF entries (see Section 4.2), we expect that poisoning would be a reality until we develop defense mechanisms or robust models on CDFs. On a more technical note, following the blueprint of this work, one can devise attacks that target the (typically simple and space-efficient) model and then apply the attack iteratively to affect the overall performance, e.g., via a greedy approach. For example, works such as the PGM INDEX [15], use linear regression as last-stage indexing model, but employs intricate model for first-stage partitioning. We believe that by applying directly our technique from Section 4 for the last stage of PGM, one can develop tailored attacks to a significant number of these variations. Other works such as Setiawan *et al.* [47] propose the use of function interpolation as a second-stage model. A possible attack may involve maximizing the MSE on *function interpolation*, as opposed to the MSE of linear regression.

**On Defending Against Poisoning in LIS.** Defending against the proposed poisoning attacks of this work is challenging. We see two ways forward, one is to swap every linear regression model to a more robust model and the other is to develop defense mechanisms that can identify poisoning points.

Regarding the model substitution approach, LIS is a more favorable option compared to traditional index data structures because of (i) the storage and (ii) time efficiency of the model used, i.e., linear regression [7, 30]. Specifically, an LIS based on linear regression requires only two or four 8-byte double values *per model* and because RMI has only 2-3 levels, it requires storing a large number of such models. Thanks to the small storage space of each model, LIS achieves a magnitude smaller memory footprint in main memory storage compared to B+Trees (have to store keys and pointers in internal nodes). Migrating a LIS to a more complex model to mitigate the risk of poisoning, would introduce significant overhead in time and/or storage, thus, substantially reduce (or even eliminate) the performance advantage of LIS over traditional index structures.

Regarding the poisoning detection approach, most poisoning defenses focus on neural networks [34, 46, 57] and classification tasks [6]; there are very few works on mitigation for poisoning of linear regression. Contrary to traditional poisoning attacks that tend to introduce outliers, poisoning CDF functions tends to populate relatively dense areas of the key space and as a result we expect that the poisoning points (in this new CDF context) are hard to identify. Jagielski *et al.* [26] proposed a poisoning detection algorithm, TRIM, on linear regressions. TRIM searches for the keys that cause the largest loss and labels them as poisoning keys. There are two major limitations in applying TRIM to our setting. Firstly, in our setting, the rank of each key depends on the value of all other keys in the dataset; this implies that TRIM has to iteratively re-calibrate its parameters and as a result become extremely inefficient. Secondly, our poisoning keys are typically concentrated around legitimate keys which makes it hard to detect. Thus, we believe that TRIM cannot remove poisoning keys without removing a significant number of legitimate keys.

Overall, the intricacies of the CDF itself, i.e., each update affects multiple entries, as well as the behavior of poisoning algorithms in this context, i.e., populate dense areas instead of sparse, suggest that we need to develop new and tailored mitigations for CDFs.

## 7 CONCLUSION

Learned index structures [30] aim at achieving the functionality of data structures using machine learning models. What differentiates the LIS paradigm from previous approaches is that ML-models adapt to the data at hand. In this work, we propose data poisoning attacks that exploit the above advantage for adversarial purposes. Our attacks poison ML-models on CDFs, which is a family of functions that has not been studied under an adversarial lens. We demonstrate our attacks on synthetic and real-world datasets under various distributions and LIS parameterizations and show that they achieve significant slow-down in every tested scenario.

# A APPENDIX
## A.1 Additional Illustration of the Proposed Poisoning Attack on CDFs

We show in Figure 8 an application of the poisoning attack on CDF using 10 poisoning keys on a data set of 90 uniformly distributed keys. Our attack increases the error by 7.4×. Each point contributes to the overall error by its distance to the regression line, indicated with blue vertical lines. Poisoning keys, colored red, are clustered on dense areas so as to exacerbate the non-linearity of the poisoned CDF. Intuitively, the attack creates non-linearity by "overpopulating" the dense area of the key values in dataset. As a result, it is hard to distinguish the poisoning keys from existing keys.

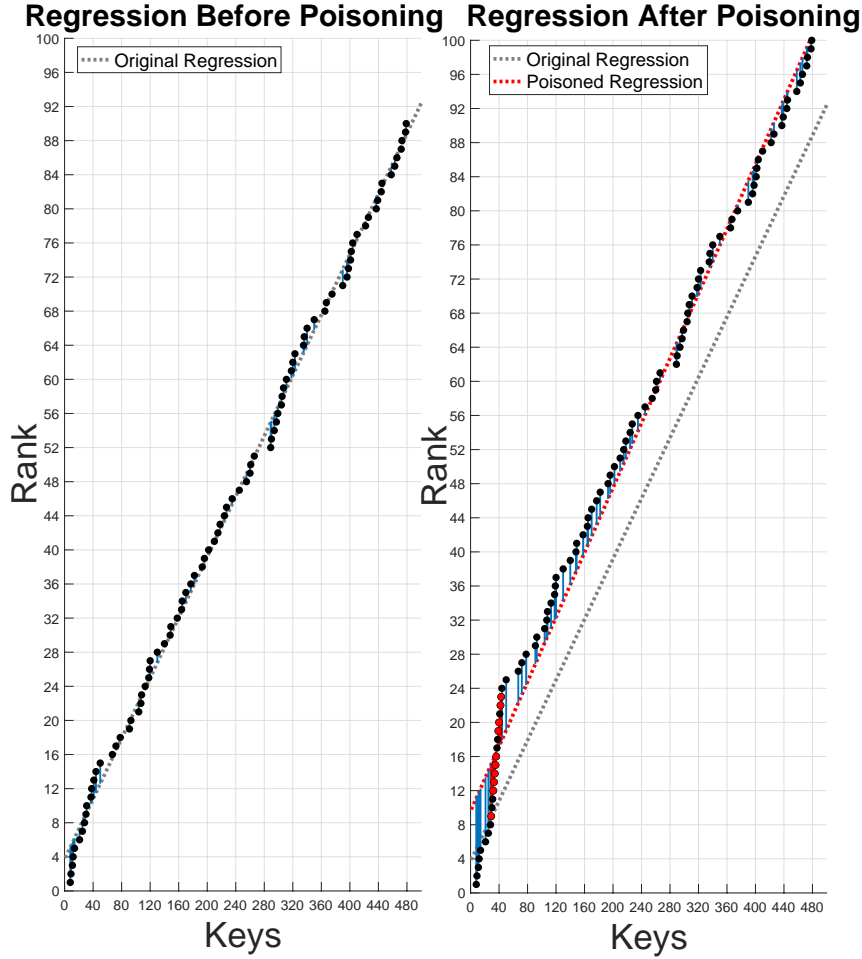

Figure 8: Example of poisoning attack using 10 poisoning keys on a data set of 90 uniformly distributed keys.

## A.2 Evidence of Optimality for Conjecture 1

Recall that the loss function $L(k_p)$ is:

$$L(k_p) = \min_{w,b}(\sum_{i=1}^{n+1}(wk_i + b - r_i)^2) = \min_{w,b}(\sum_{i \neq p}((wk_i + b - r_i)^2) + (wk_p + b - r_p)^2)$$

The objective is finding the $k_p$ that maximizes the loss. Although $w$ and $b$ are results of minimization problem that depends on $k_p$, we denote them as variable in the following definition of Loss:

$$L'(k_p, w, b) = \sum_{i=1}^{n}(wk_i + b - r_i)^2 = \sum_{i \neq p}(wk_i + b - r_i)^2$$

To prove $L$ convex over $k_p$, it is sufficient to prove that $L'$ is convex because $L'$ is a minimization problem of $L$. For all $i \neq p$, $(wk_i + b - r_i)^2$ is a quadratic function and is thus convex. Therefore, only $(wk_p + b - r_p)^2$ is potentially non-convex ($w$ and $k_p$ are both variables). While this term is not strictly convex for any $\{w, k_p, b\}$, we observe from experiment that the non-convexity of this term does not overwhelm the rest $n - 1$ convex terms. Consider:

$$L'(k_p, w, b) + L'(k_p + 2, w + 2\epsilon_w, b + 2\epsilon_b) - 2L'(k_p + 1, w + \epsilon_w, b + \epsilon_b),$$

where $\epsilon_w$ and $\epsilon_b$ are small gradient values. After simplifying we have:

$$(\sum_{i \neq p} 2(\epsilon_w k_i + \epsilon_b)^2) + 4\epsilon_w((w + \epsilon_w)(k_p + 1) + (b + \epsilon_b) - r_p) + 6\epsilon_w^2$$

Since $\epsilon_w$ is a small value and $(w + \epsilon_w)(k_p + 1) + (b + \epsilon_b) - r_p$ the loss of poisoning key bounded by $poly(L)$, the second term is overpowered by the squared positive terms. Therefore, the loss is very likely convex.

## A.3 Evaluation of Max Memory Offset of Uniformly Distributed Data (Regression)

**Setup.** The setup is identical to the experiment in Section 4. Instead of measuring the average memory offset for each dataset, we measure the *maximum* memory offset over all queries on the keys that present in the dataset.

**Results.** Similar to average memory offset, maximum memory offset grows linearly with the size of the dataset as shown in Figure 9. In comparison, the maximum memory offset is roughly two times the average memory offset given the same setup, which implies that the skew of the distribution of memory offsets is close to center.

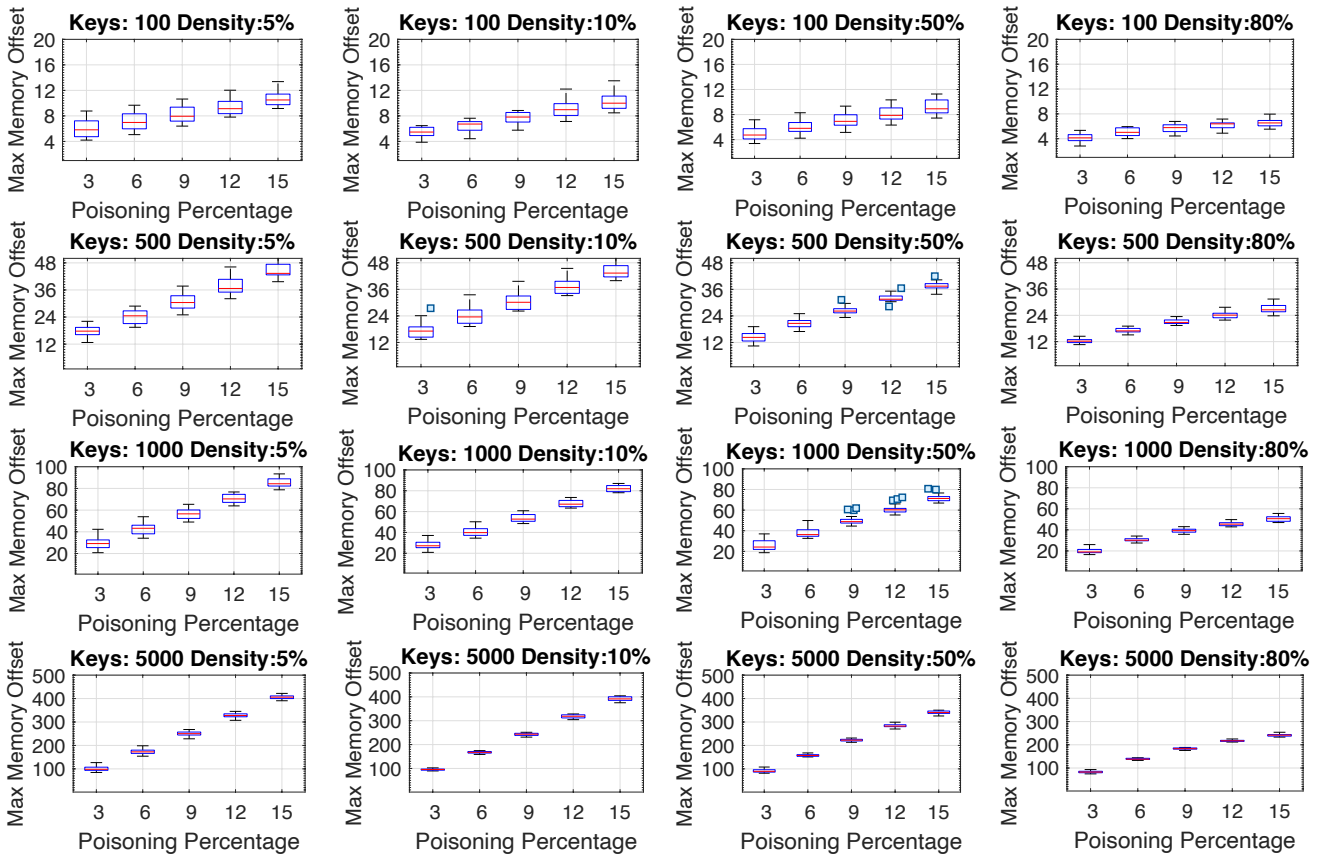

Figure 9: Evaluation of the multi-point poisoning for linear regression on CDF. Each boxplot shows the maximum model prediction memory offset of legitimate key queries over 20 distinct keysets. The legitimate keys have normal distribution. The number of legitimate keys, denoted as "Keys" and ratio between the number of legitimate keys and the size of the key domain, denoted as "Density" are presented on the title of each plot. The number of poisoning keys varies on the $X$-axis and is represented as a poisoning percentage with respect to the number of legitimate keys.

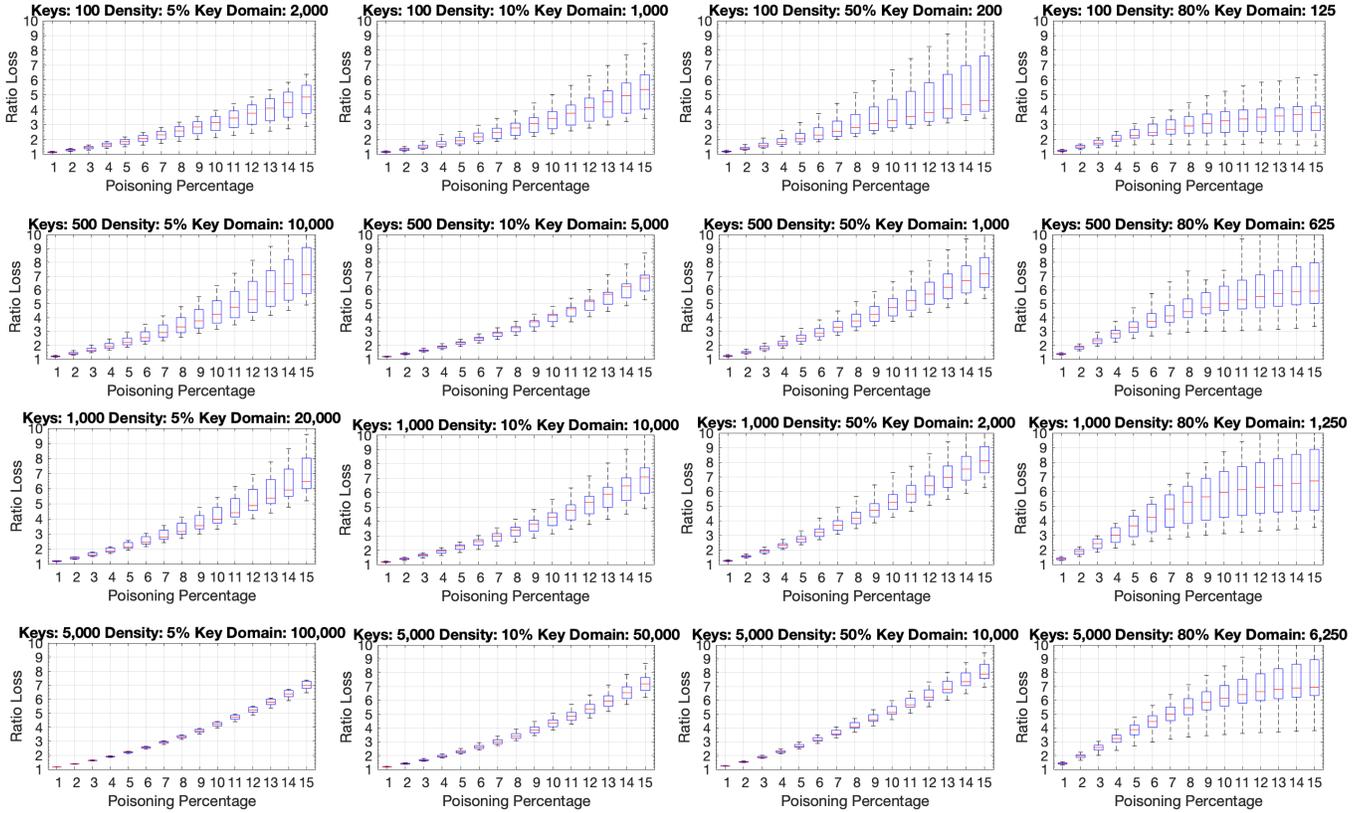

Figure 10: Evaluation of the multi-point poisoning for linear regression on CDF. Each boxplot shows the ratio of the evaluation of the MSE loss on the poisoned keyset over the evaluation of the MSE loss on the legitimate keyset over 20 runs. The legitimate keys are in normal distribution. The number of poisoning keys varies on the $X$-axis and is represented as a poisoning percentage with respect to the number of legitimate keys.

### A.4 Experiments on Synthetic Data from a Normal Distribution (Regression)

**Setup.** The setup is identical to the experiment in Section 4 but on a normal distributed dataset. Specifically, for a key domain $U = [\alpha, \beta]$ we form the normal distribution with mean $\mu = \frac{\beta+\alpha}{2}$ and standard deviation $\sigma = \frac{\beta-\alpha}{3}$.

**Results.** The results are presented in Figure 10. Because the normal distributions are not captured well by linear models, the MSE loss of the original dataset is high. Despite that, our attack achieves up to 8× increase of error.

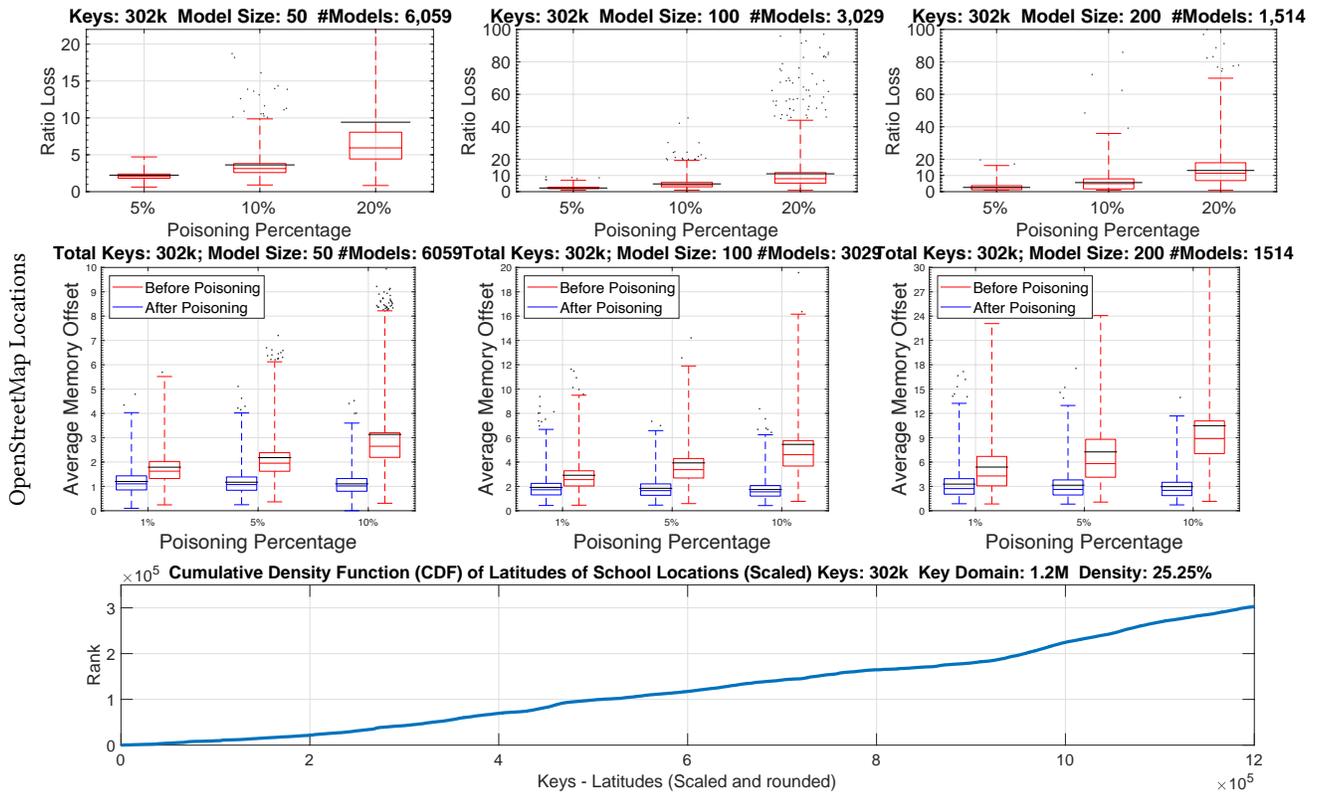

Figure 11: Evaluation of the multi-point poisoning for RMI applied on the CDF of latitudes of schools in OpenStreetMap, filtered by latitudes between -30 and +50. The $X$-axis represents different overall poisoning percentage where the second-stage poisoning threshold $\alpha$ takes value $\alpha = 3$. The third row presents the CDF.

## A.5 Evaluation on Real-World Data: OpenStreetMap Geolocations

**Setup.** We use the publicly available geolocation dataset [44] and index by the latitudes of locations. We pick the data points that are labelled as schools over the world with latitudes between -30 and +50; and scale up the latitudes by 15,000 before rounding to achieve uniqueness of keys. The final dataset has $n = 302,973$ keys in a key universe of $m = 1,200,000$, which yields a 25% density. The cumulative distribution is depicted in Figure 11.

In this experiment, we test three different RMI model setups. In the first setup, we initialize the second-stage models with 50 keys each, which translates to 6,059 second-stage models. In the second setup, we initialize the second-stage with 100 keys each, which translates to 3,029 models. In the third setup, we initialize the second-stage with 200 keys each, which translates to 1,514 models. For all the setups, we consider the poisoning threshold of the second stage model to have parameter $\alpha = 3$. Finally, we considered poisoning percentages 5%, 10%, and 20%.

**Results.** The results of our evaluation are presented in Figure 11. For a fixed setup, we observe that as the poisoning percentage increases, the ratio loss increases as well. Among all experiments in both datasets, the RMI loss increases between 4× to 24×. Since the number of poisoning keys per second-stage model is a percentage over its number of keys, we see that larger models allow more poisoning and, consequently, larger RMI error. The discrepancy between the effectiveness of the attack on synthetic data and the real data is partly explained by the fact that the synthetic dataset is four orders of magnitude larger than the real dataset.